\documentclass[final,5p,twocolumn,times,sort&compress]{elsarticle}

\usepackage{graphicx}
\usepackage{graphicx}
\usepackage{amsmath,bm}
\usepackage{flushend}
\usepackage{color}

\usepackage{soul,xcolor}
\setstcolor{green}

\def\al{\alpha}
\def\be{\beta}
\def\ga{\gamma}
\def\de{\delta}
\def\ep{\epsilon}

\def\ze{\zeta}

\def\ka{\kappa}
\def\la{\lambda}

\def\rh{\rho}

\def\si{\sigma}

\def\ps{\psi}

\def\Ga{\Gamma}

\def\mn{{\mu\nu}}

\def\prt{\partial}

\def\fr#1#2{{{#1} \over {#2}}}
\def\half{{\textstyle{1\over 2}}}
\def\lsim{\mathrel{\rlap{\lower4pt\hbox{\hskip1pt$\sim$}}
    \raise1pt\hbox{$<$}}}
\def\gsim{\mathrel{\rlap{\lower4pt\hbox{\hskip1pt$\sim$}}
    \raise1pt\hbox{$>$}}}
\def\sqr#1#2{{\vcenter{\vbox{\hrule height.#2pt
         \hbox{\vrule width.#2pt height#1pt \kern#1pt
         \vrule width.#2pt}
         \hrule height.#2pt}}}}
\newcommand{\beq}{\begin{equation}}
\newcommand{\eeq}{\end{equation}}
\newcommand{\bea}{\begin{eqnarray}}
\newcommand{\eea}{\end{eqnarray}}
\newcommand{\rf}[1]{(\ref{#1})}
\newcommand{\nn}{\nonumber}

\def\etal{{\it et al.}}

\def\cL{{\cal L}}
\def\half{{\textstyle{1\over 2}}}
\def\quar{{\textstyle{1\over 4}}}
\def\eigh{{\textstyle{1\over 8}}}

\def\lsim{\mathrel{\rlap{\lower4pt\hbox{\hskip1pt$\sim$}}
    \raise1pt\hbox{$<$}}}
\def\gsim{\mathrel{\rlap{\lower4pt\hbox{\hskip1pt$\sim$}}
    \raise1pt\hbox{$>$}}}

\def\prt{\partial}
\def\lrprt#1{\hskip-2pt\stackrel{\leftrightarrow}{\prt_{#1}}\hskip-2pt}

\def\etal{{\it et al.}}

\def\mn{{\mu\nu}}

\def\psb{\overline{\ps}{}}

\def\nm{N}
\def\nmetl#1#2#3{\nm_{{#1}{#2}{#3}}}
\def\nmo{{(\nm_1)}{}}
\def\nmt{{(\nm_2)}{}}
\def\sym#1#2#3{S_{{#1}{#2}{#3}}}
\def\symluu#1#2#3{S_{#1}{}^{#2}{}^{#3}}
\def\mix#1#2#3{M_{#1}{}_{#2}{}_{#3}}
\def\mixllu#1#2#3{M_{#1}{}_{#2}{}^{#3}}
\def\mixuuu#1#2#3{M^{#1}{}^{#2}{}^{#3}}

\def\xx#1#2{\ze^{(#1)}_{#2}}

\begin{document}

\begin{frontmatter}

\title{Constraining Spacetime Nonmetricity
with Neutron Spin Rotation in Liquid ${}^4$He}

\author[IUCSS,Hanover]{Ralf Lehnert}
\author[IUCSS,IU,CEEM]{W.M.~Snow}
\author[IUCSS,China]{Zhi Xiao}
\author[IU]{Rui Xu}

\address[IUCSS]{Indiana University Center for Spacetime Symmetries, Bloomington, IN 47405, USA}
\address[Hanover]{{Leibniz Universit\"at Hannover, Welfengarten 1, 30167 Hannover, Germany}}
\address[IU]{Physics Department, Indiana University, Bloomington, IN 47405, USA}
\address[CEEM]{Center for Exploration of Energy and Matter,
Indiana University,\\
Bloomington, IN 47408, USA}
\address[China]{Department of Mathematics and Physics, North China Electric Power University, Beijing 102206, China}

\begin{abstract}
General spacetime nonmetricity coupled to neutrons is studied. 
In this context, 
it is shown that 
certain nonmetricity components 
can generate a rotation of the neutron's spin. 
Available data on this effect 
obtained from slow-neutron propagation 
in liquid helium 
are used to constrain isotropic nonmetricity components
at the level of $10^{-22}\,$GeV. 
These results represent the first limit 
on the nonmetricity $\zeta^{(6)}_2S_{000}$ parameter 
as well as the first measurement of nonmetricity 
inside matter.

\end{abstract}
\end{frontmatter}


\section{Introduction}

The idea that 
spacetime geometry represents a dynamical physical entity 
has been remarkably successful 
in the description of classical gravitational phenomena. 
For example,
General Relativity, 
which is based on Riemannian geometry,
has recently passed a further experimental test:
the theory predicts gravitational waves, 
and these have indeed been observed 
by the LIGO Scientific Collaboration 
and the Virgo Collaboration~\cite{gravitywaves}.

At the same time,
a number of observational as well as theoretical issues
motivate the construction and study of 
alternative gravity theories.
Most of these efforts 
recognize the elegance and success of 
a geometric underpinning for gravitational phenomena
and therefore retain this feature in model building.
One popular approach in this context, 
known as metric-affine gravity~\cite{bl13},
employs an underlying geometry 
more general than that of a Riemannian manifold. 
The basic idea behind this approach 
can be summarized as relaxing
the metric-compatibility condition $D_{\al}g_{\be\ga}=0$
and the symmetry condition on the connection coefficients
$\Ga^\al{}_{\be\ga}-\Ga^\al{}_{\ga\be}=0$.
In general, 
this idea introduces 
two tensor fields
\beq
\label{NMandTdef}
N_{\al\be\ga}\equiv -D_{\al}g_{\be\ga}\,,\qquad
T^\al{}_{\be\ga}\equiv \Ga^\al{}_{\be\ga}-\Ga^\al{}_{\ga\be}\,,
\eeq
relative to the Riemannian case 
known as nonmetricity and torsion,
respectively.

The specialized situation 
in which the nonmetricity tensor vanishes $N_{\al\be\ga}=0$ 
and only torsion is nonzero
represents the widely known Einstein--Cartan theory~\cite{ca22}.
In that context, 
torsion has been the subject of various investigations
during the last four decades~\cite{torsionreviews}. 
Considering the question of the presence of torsion in nature
as an experimental one 
has spawned numerous phenomenological studies of torsion~\cite{Overview,micro,SunSource,astro,Kaons,CL97,LHC,GravWave,GProbeB,TorsionSME,PrefFrTests} yielding bounds on various torsion couplings.

An analogous phenomenological investigation of nonmetricity 
has been instigated last year~\cite{nm}.
Paralleling the torsion case,
that analysis treats the question 
regarding the presence of nonmetricity 
as an experimental one,
and the nonmetricity field $N_{\al\be\ga}$ is taken 
as a large-scale background 
extending across the solar system. 
The particular physical situation 
considered in Ref.~\cite{nm} 
lends itself to an effective-field-theory description 
in which $N_{\al\be\ga}$ represents 
a prescribed external field 
selecting preferred spacetime directions. 
Thus, 
such a set-up embodies in essence
a Lorentz-violating scenario 
amenable to theoretical treatment
via the Standard-Model Extension (SME) framework~\cite{sme}. 
For example, 
sidereal and annual variations of physical observables 
resulting from the motion of an Earth-based laboratory 
through this solar-system nonmetricity background
represent a class of characteristic experimental signals
in that context~\cite{DataTables}.

The present work
employs a similar idea
to obtain further, complementary constraints on nonmetricity.
The specific set-up we have in mind 
consists of liquid ${}^4$He 
as the nonmetricity source. 
Polarized neutrons generated at
the slow-neutron beamline 
at the National Institute of Standards and Technology (NIST) Center for Neutron Research 
traverse the helium 
and serve as the nonmetricity probe.
It is apparent that 
our set-up involves an Earth-based nonmetricity probe.
Thus, 
the key difference between our study 
and that in Ref.~\cite{nm} is that
we examine the situation of nonmetricity 
sourced locally in a terrestrial laboratory by the ${}^4$He. 
This implies that 
the presumed nonmetricity 
in our case is comoving with the laboratory, 
and thus the neutron probe, 
which precludes certain experimental signatures, 
such as sidereal and annual variations.
Instead,
we utilize the prediction presented below that
certain components of $N_{\al\beta\ga}$ 
lead to neutron spin rotation 
in this system.

The outline of this paper is as follows.
Section~\ref{theory} 
reviews the basic ideas 
behind the effective-field-theory description 
of a background $N_{\al\be\ga}$ 
in flat Minkowski space 
and derives the resulting spin motion for nonrelativistic neutrons.
This effect provides the basis for 
our limits on nonmetricity.
The details of the measurement of 
neutron spin rotation in liquid ${}^4$He 
including the experimental set-up 
are discussed in Sec.~\ref{exp}.
A brief summary 
is contained in Sec.~\ref{sum}. 
Throughout,
we adopt natural units 
$c=\hbar=1$. 
Our conventions for the metric signature
and the Levi--Civita symbol are
$\eta^{\mn}=\textrm{diag}(+,-,-,-)$
and $\epsilon^{0123}=+1$, 
respectively.

\section{Theory}
\label{theory}

Our analysis is based on 
the approach to nonmetricity couplings 
taken in Ref.~\cite{nm}, 
so we begin with a brief review 
of that approach.
The basic idea is to follow the usual reasoning 
that the construction of an effective Lagrangian 
should include all terms 
compatible with the symmetries of the model.
In the present context, 
possible couplings between the background nonmetricity $N_{\al\be\ga}$ 
and the polarized-neutron probe 
need to be classified.
Since we are interested in a low-energy experiment,
we may disregard the neutron's internal structure 
and model it as a point Dirac fermion 
with free Lagrangian 
${\cal L}_0=\half\, \psb \, \ga^\mu i\!\lrprt\mu \ps - m\, \psb \ps $, 
where $m$ denotes the neutron mass. 
Conventional gravitational effects are negligible, 
so that the flat-spacetime Minkowski limit $g^\mn\to\eta^\mn$ 
suffices for our present purposes.

The next step is to enumerate 
possible couplings of $\ps$ 
to the background nonmetricity $N_{\al\be\ga}$. 
This yields a hierarchy of possible Lagrangian terms ${\cal L}_N^{(n)}$ 
labeled by the mass dimension $n$ 
of the corresponding field operator:
\beq
\cL_N
=
\cL_0
+\cL_N^{(4)}
+\cL_N^{(5)}
+\cL_N^{(6)}
+ \ldots\,.
\label{lagexp}
\eeq
For the experimental set-up we have in mind, 
nonmetricity couplings affecting the propagation of neutrons 
are the most relevant ones. 
Moreover, 
$N_{\al\be\ga}$ must be small on observational grounds. 
We therefore focus on contributions to ${\cal L}_N^{(n)}$ that
are quadratic in $\ps$ and linear in $N_{\al\be\ga}$.
General arguments in effective field theory suggest that
Lagrangian terms of lower mass dimension $n$
may be more dominant. 
Capturing the leading effects of 
all nonmetricity components 
then requires inclusion of Lagrangian terms
up to mass dimension $n=6$~\cite{nm}.

The construction of the explicit form of 
each individual contribution ${\cal L}_N^{(n)}$ 
is most easily achieved by 
decomposing $N_{\al\be\ga}$ 
into its Lorentz-irreducible pieces. 
These are given by two vectors $(N_1)_{\mu}$ and $(N_2)_{\mu}$,
a totally symmetric rank-three tensor $S_{\mu\al\be}$, 
and a rank-three tensor $M_{\mu\al\be}$ 
with mixed symmetry~\cite{nm}:
\bea
\label{irrep}
(N_1)_\mu 
\hskip -7pt 
&\equiv& 
\hskip -7pt 
-\eta^{\al\be} \nmetl \mu\al\be\, ,
\nn\\[7pt]
(N_2)_\mu 
\hskip -7pt
&\equiv& 
\hskip -7pt
-\eta^{\al\be} 
\nmetl \al\mu\be\, ,
\nn\\[7pt]
\sym \mu\al\be 
\hskip -7pt 
&\equiv& 
\hskip -7pt
\tfrac 1 3 \big[ \nmetl \mu\al\be + \nmetl \al\be\mu + \nmetl \be\mu\al \big] 
\nn\\
&& 
\hskip -7pt 
+ \tfrac 1 {18} 
\big[ \nmo_\mu \,\eta_{\al\be} + \nmo_\al \,\eta_{\be\mu} + \nmo_\be \,\eta_{\mu\al} \big] 
\nn\\
&& 
\hskip -7pt 
+ \tfrac 1 {9} 
\big[ \nmt_\mu \,\eta_{\al\be} + \nmt_\al \,\eta_{\be\mu} + \nmt_\be \,\eta_{\mu\al} \big]\, ,
\nn\\[7pt]
\mix \mu\al\be
\hskip -7pt 
&\equiv& 
\hskip -7pt
\tfrac 1 3 \big[
2 \nmetl \mu\al\be - \nmetl \al\be\mu - \nmetl \be\mu\al
\big]
\nn\\
&& 
\hskip -7pt 
+ \tfrac 1 9 \big[
2 \nmo_\mu \,\eta_{\al\be} - \nmo_\al \,\eta_{\be\mu} - \nmo_\be \,\eta_{\al\mu}
\big]
\nn\\
&& 
\hskip -7pt 
- \tfrac 1 9 \big[
2 \nmt_\mu \,\eta_{\al\be} - \nmt_\al \,\eta_{\be\mu} - \nmt_\be \,\eta_{\al\mu}
\big]\,.
\quad
\eea
With these pieces,
the nonmetricty tensor can be reconstructed as follows~\cite{nm}:
\bea
\label{recon}
\nmetl \mu\al\be \hskip -7pt &=& \hskip -7pt
\tfrac 1{18} \big[
- 5 \nmo_\mu \eta_{\al\be} + \nmo_\al \eta_{\be\mu} + \nmo_\be \eta_{\mu\al}
\nn\\
&& \hskip -7pt
+ 2 \nmt_\mu \eta_{\al\be} - 4 \nmt_\al \eta_{\be\mu} - 4 \nmt_\be \eta_{\mu\al}\big]
\nn\\
&& \hskip -7pt
+ \sym \mu\al\be + \mix \mu\al\be\,.
\eea
The sign changes in Eqs.~\rf{irrep}, \rf{recon},
and some subsequent equations
relative to the corresponding equations in Ref.~\cite{nm} 
arise due to differing conventions 
for the metric signature 
and for the sign of the Levi--Civita symbol.
We also remark that 
although Eqs.~\rf{irrep} and~\rf{recon} 
employ a notation similar to that 
for the irreducible components of 
torsion $T^{\al}{}_{\be\ga}$~\cite{TorsionSME},
the nonmentricity and torsion pieces 
are unrelated.

With this decomposition, 
the following 
Lagrangian contributions can be constructed~\cite{nm}:
\bea
\label{lagr}
\hskip -10pt \cL_\nm^{(4)}
\hskip -7pt &=& \hskip -7pt
\ze^{(4)}_1\, \nmo_\mu\, \psb\ga^\mu\ps
+ \ze^{(4)}_2\, \nmo_\mu\, \psb\ga_5\ga^\mu\ps
\nn\\ &&
+ \ze^{(4)}_3\, \nmt_\mu\, \psb\ga^\mu\ps
+ \ze^{(4)}_4\, \nmt_\mu\, \psb\ga_5\ga^\mu\ps\,,
\nn\\[7pt]
\hskip -10pt \cL_\nm^{(5)}
\hskip -7pt &=& \hskip -7pt
-\half i \ze^{(5)}_1\, \nmo^\mu\, \psb\lrprt\mu \ps
-\half \ze^{(5)}_2\, \nmo^\mu\, \psb \ga_5 \lrprt\mu \ps
\nn\\ &&
- \half i \ze^{(5)}_3\, \nmt^\mu\, \psb\lrprt\mu \ps
- \half \ze^{(5)}_4\, \nmt^\mu\, \psb \ga_5\lrprt\mu\ps
\nn\\ &&
- \quar i \ze^{(5)}_5\, \mixllu \mu\nu\rh\, \psb \si^\mn \lrprt\rh \ps 
\nn\\ &&
+ \eigh i \ze^{(5)}_6\, \ep_{\ka\la\mu\nu}\, 
\mixuuu \ka\la\rh\, \psb \si^\mn \lrprt\rh \ps 
\nn\\ &&
+ \half i \ze^{(5)}_7\, \nmo_\mu\, \psb \si^\mn \lrprt\nu \ps
+ \half i \ze^{(5)}_8\, \nmt_\mu\, \psb \si^\mn \lrprt\nu \ps
\nn\\ &&
- \quar i \ze^{(5)}_9\, \ep^{\la\mu\nu\rh} 
\nmo_\la\, \psb \si_\mn \lrprt\rh \ps  
\nn\\ &&
- \quar i \ze^{(5)}_{10}\, \ep^{\la\mu\nu\rh} 
\nmt_\la\, \psb \si_\mn \lrprt\rh \ps\,,
\nn\\[10pt]
\hskip -10pt \cL_\nm^{(6)}
\hskip -7pt &\supset& \hskip -7pt
- \quar \ze^{(6)}_1\, 
\symluu \la\mu\nu\, \psb\ga^\la \prt_\mu \prt_\nu \ps
+ {\rm h.c.}
\nn\\
&&
- \quar \ze^{(6)}_2\, 
\symluu \la\mu\nu\, \psb\ga_5\ga^\la \prt_\mu \prt_\nu \ps
+ {\rm h.c.}
\eea
Here, 
the real-valued couplings $\zeta^{(n)}_l$ 
are taken as free parameters; 
they can in principle be fixed by 
specifying a definite underlying nonmetricity model.
For the mass-dimension six term ${\cal L}^{(6)}_N$,
we have only listed those contributions that
contain the $S_{\mu\al\be}$ irreducible piece;
all other components of $N_{\al\be\ga}$ are already present 
in the terms ${\cal L}^{(4)}_N$ or ${\cal L}^{(5)}_N$ 
of lower mass dimension. 

Equations~\rf{lagexp}, \rf{irrep}, and~\rf{lagr} 
determine the low-energy neutron effective Lagrangian 
in the presence of general background nonmetricity
relevant for the experimental situation 
we have in mind.  
We note, 
however, 
that the terms~(\ref{lagr}) 
would generally be viewed as part of a more complete 
Lagrangian ${\cal L}\supset{\cal L}_N$
that also treats $N_{\al\be\ga}$ as a dynamical variable. 
The nonmetricity field equations 
then contain $\partial {\cal L}/\partial N_{\al\be\ga}$, 
and thus neutron source terms. 
This idea provides the justification 
for taking the ${}^4$He nucleus 
as a nonmetricity source 
in the experimental set-up discussed below.
The protons and electrons of the ${}^4$He atom 
may produce additional nonmetricity contributions 
if these particles exhibit nonmetricty couplings 
analogous to those in Eq.~(\ref{lagr}). 
In what follows, 
we make no assumptions regarding the dynamics of $N_{\al\be\ga}$ 
or additional nonmetricity--matter couplings;
we simply presume that 
the ${}^4$He generates some nonzero nonmetricity. 

A model refinement can be achieved by
focusing on the leading contribution to $N_{\al\be\ga}$.  
Note that $N_{\al\be\ga}=N_{\al\be\ga}(x)$
must exhibit a nontrivial spacetime dependence
determined by the interatomic distance 
and the velocity of the ${}^4$He atoms. 
However, 
the random nature of these two quantities suggests that 
the leading nonmetricity effects 
are actually governed by 
the spacetime average $\langle N_{\al\be\ga}(x)\rangle$. 
For this reason,
we may take $N_{\al\be\ga}=\textrm{const}$.\ in what follows.
The nonmetricity contributions~\rf{lagr} 
then form a subset of the flat-space SME Lagrangian, 
a fact that permits us
to employ the full repertoire of theoretical tools 
developed for the SME framework. 

One such SME result relevant for the present situation 
concerns the observability of constant background fields~\cite{sme,redefs}.
For example, 
it is known that contributions 
associated with the couplings 
$\ze^{(4)}_1$, 
$\ze^{(4)}_3$, 
$\ze^{(5)}_1$, 
$\ze^{(5)}_2$, 
$\ze^{(5)}_3$, 
$\ze^{(5)}_4$,
$\ze^{(5)}_7$,
and $ \ze^{(5)}_8$
can be removed from the Lagrangian---at least at linear order---via judiciously chosen field redefinitions. 
We may therefore disregard these terms in what follows.
Their measurement would require situations 
involving nonconstant $N_{\al\be\ga}$, 
the presence of gravity, 
or the consideration of higher-order effects.

An additional simplification arises from
the isotropy of the liquid helium. 
The ${}^4$He ground state has spin zero,
so anisotropies would have to be tied to
excited states of ${}^4$He or 
arrangements of the helium atoms 
involving preferred directions.
However, 
the absence of polarization 
and the aforementioned random nature 
of both position and velocity of individual ${}^4$He particles
precludes sizeable, large-scale anisotropies. 
The leading background nonmetricty contributions
generated by the liquid-helium bath 
can therefore also be taken as isotropic 
in the helium's center-of-mass frame. 
It follows that 
the present experimental set-up
is only sensitive to the rotationally invariant pieces of $N_{\al\be\ga}$.

To uncover the isotropic content of $N_{\al\be\ga}$,
we may proceed by inspecting its irreducible pieces~\rf{irrep}. 
Clearly,
components without spatial indices are rotation symmetric: 
$\nmo_0$, $\nmt_0$, and $S_{0 0 0}$. 
Note that $M_{\al\be\ga}$ obeys the cyclic property 
\beq
\label{cyclic}
M_{\al\be\ga}+M_{\be\ga\al}+M_{\ga\al\be}=0\,,
\eeq
which implies $M_{000}=0$.
Further isotropic components in $S$ and $M$ with spatial indices 
must have spatial-index structure $\delta_{jk}$ or $\epsilon_{jkl}$, 
where Latin indices run from 1 to 3.
Since both $S$ and $M$ are symmetric in their last two indices, 
they cannot contain pieces of $\epsilon_{jkl}$. 
This only leaves contributions involving $\delta_{jk}$.
But these do not yield independent isotropic contributions 
because both $S$ and $M$ are traceless.
To see this,
consider as an example 
a piece of the form $S_{0jk}=s\, \delta _{jk}$, 
where $s$ is the isotropy parameter in question.
But $S$ is traceless, 
so that we have 
$0=
S_{0\al\be}\,\eta^{\al\be}=
S_{000}-S_{0jk}\,\delta_{jk}=
S_{000}-s\, \delta_{jk}\,\delta_{jk}$.
It follows that
$3 s= S_{000}$
does not represent 
an additional independent isotropic contribution to $S$.
An analogous reasoning applies to $M$,
so that $\nmo_0$, $\nmt_0$, and $S_{0 0 0}$ 
are indeed the only isotropic nonmetricity components.

The model determined by Eqs.~\rf{lagexp} and~\rf{lagr} 
permits a fully relativistic description 
of all dominant nonmetricity effects  
on the propagation of both neutrons and antineutrons
in the present context. 
Since our current goal is an analysis of 
the spin motion of slow neutrons,
we may disregard all antineutron physics, 
and focus entirely on 
the $2\times 2$ nonrelativistic 
neutron Hamiltonian $h=h_0+\delta h +\delta h_s$ 
resulting from our model Lagrangian~\rf{lagr}.
Here,
$h_0$ is the ordinary nonrelativistic piece.
The spin-independent nonmetricty contribution $\delta h$ 
is irrelevant for this work. 
The spin-dependent correction $\delta h_s$ 
resulting from Eq.~\rf{lagr} 
can be gleaned from previously established SME studies~\cite{nonrel_ham}.
The result for both isotropic as well as anisotropic contribution reads
\bea
\label{ham}
\hskip -10pt
 \de h_s 
=&&
\hskip -15pt
\Big[ \Big(\ze^{(4)}_2 - m\, \ze^{(5)}_9\Big) \nmo_{j}
 + \Big(\ze^{(4)}_4 - m\, \ze^{(5)}_{10}\Big) \nmt_{j} \Big]\, \sigma^j
\nn\\ &&
\hskip -40pt
 +\Big[ \Big(\ze^{(4)}_2 - m\, \ze^{(5)}_9\Big) \nmo_{0} 
 + \Big(\ze^{(4)}_4 - m\, \ze^{(5)}_{10}\Big) \nmt_{0} \Big]\,
 \fr {\vec p \cdot \vec \si}{m} 
\nn\\ &&
\hskip -40pt
+\half \Big[ \ze^{(5)}_5 \tilde{M}_{j\al\be}
+\tfrac{3}{2}\,\ze^{(5)}_6 M_{j\al\be}
+ m\, \ze^{(6)}_2 S_{j\al\be}\Big]\, 
\frac{p^\al p^\be \si^j}{m} 
\nn\\ &&
\hskip -40pt
+ \half\ze^{(6)}_2 S_{0\al\be}\,  \fr{p^\al p^\be \vec p \cdot \vec \si}{m}\,.
\eea
This expression contains the leading contribution 
in the nonrelativistic order $|\vec{p}|/m$
for each nonmetricity component. 
In the above equation,
we have set 
$\tilde{M}_{\al\be\ga}\equiv \epsilon_{\al\be}{}^{\mu\nu}M_{\mu\nu\ga}$.
Moreover, 
$p^\mu = (p^0, \vec{p})=(p^0,p^j)$ 
denotes the neutron's 4-momentum, 
and $\sigma^j$
are the usual Pauli matrices. 
Note that nonmetricity effects 
corresponding to $\zeta^{(6)}_1$  
only produce spin-independent effects.
They are therefore absent from $\delta h_s$ 
and cannot be determined 
by observations of neutron spin rotation.

\section{Experimental Analysis}
\label{exp}

To extract experimental signatures 
resulting from the nonmetricty correction~\rf{ham}, 
we analyze the aforementioned experimental situation,
namely spin motion of a neutron 
as it passes through liquid ${}^4$He. 
As argued above, 
our Lagrangian~(\ref{lagr}) implies that neutrons,
and hence ${}^4$He nuclei, 
can generate nonmetricity. 
The injected neutron beam would then be affected 
by this nonmetricity background. 
Moreover,
our ``in-matter" approach permits us to search for
short-ranged or non-propagating nonmetricity. 
In particular,
this encompasses situations 
analogous to minimally coupled torsion,
where the torsion tensor vanishes 
outside the spin-density source~\cite{torsionreviews}.
Such an approach rests on the premise 
that the probe penetrates the matter 
and that the effects of 
conventional Standard-Model (SM) physics 
are minimized. 
The ${}^4$He--neutron system 
appears to be ideal in this respect 
for two reasons.
First,
the neutron mean free path inside liquid ${}^4$He 
is relatively long 
allowing for the accumulation of 
the predicted spin-rotation effect. 
This is due to the small elastic 
and the essentially vanishing inelastic cross sections 
as well as rapidly decreasing neutron--phonon scattering 
as $T\to 0$.
Second,
contamination of the nonmetricity spin rotation
by ordinary SM physics
can be excluded on the grounds that 
these conventional effects 
lie below the current detection sensitivity.
This latter fact is explained in more detail below.

The rotation of the spin 
of a transversely polarized slow-neutron beam 
is called neutron optical activity.  
It is quantified by the rotary power 
$d\phi_{PV}/dL$ 
defined as the rotation angle $\phi_{PV}$ of the neutron spin 
about the neutron momentum $\vec{p}$
per traversed distance $L$.
The nonmetricity correction~\rf{ham} 
leads to the following expression 
for the rotary power:
\bea
\label{spinrot}
\frac{d\phi_{PV}}{dL}
\hskip -7pt 
&=& 
\hskip -7pt 
2 \Big(\ze^{(4)}_2 - m\, \ze^{(5)}_9\Big) \nmo_{0} + 2 \Big(\ze^{(4)}_4 - m\, \ze^{(5)}_{10}\Big) \nmt_{0} 
\nn\\ &&
\hskip -7pt 
 + m^2\ze^{(6)}_2 S_{000}\,,
\eea
where we have implemented the isotropic limit. 
The neutron rotary power 
is amenable to high-precision experimental studies 
and can therefore be employed 
to measure or constrain 
the combination of nonmetricity components 
appearing on the right-hand side of Eq.~\rf{spinrot}. 

The experiment 
described in detail below 
measured the neutron rotary power 
to be
\beq
\label{measurement}
\frac{d\phi_{PV}}{dL}=+1.7\pm9.1\textrm{(stat.)}\pm1.4\textrm{(sys)}\times10^{-7}\,\textrm{rad/m}
\eeq
at the 1-$\sigma$ level. 
Conversion to natural units 
together with Eq.~\rf{spinrot}
yields the following  nonmetricity measurement: 
\bea
\label{nonmetr_measurem}
\hskip -15pt 2 \Big(\ze^{(4)}_2 - m\, \ze^{(5)}_9\Big) \nmo_{0} + 2 \Big(\ze^{(4)}_4 - m\, \ze^{(5)}_{10}\Big) \nmt_{0}  + m^2\,\ze^{(6)}_2 S_{000}\nn\\
=(3.4\pm 18.2)\times 10^{-23}\,\textrm{GeV}\,.
\eea
We interpret this result as the 2-$\sigma$ constraint
\bea
\label{bound1}
\hskip -20pt \Big|2 \Big(\ze^{(4)}_2 - m\, \ze^{(5)}_9\Big) \nmo_{0} + 2 \Big(\ze^{(4)}_4 - m\, \ze^{(5)}_{10}\Big) \nmt_{0}  + m^2\,\ze^{(6)}_2 S_{000}\Big|
\nn\\
\hspace{-100pt} <3.6 \times 10^{-22} \,\textrm{GeV}\,.
\eea
Disregarding the possibility of 
extremely fine-tuned cancellations 
between the various nonmetricity couplings in the constraint~\rf{bound1},
we can estimate the following individual bounds: 
\bea
\label{bound2}
|\xx	4	2	\nmo_0| < 10^{-22}\,\textrm{GeV}\,, && 
|\xx	4	4	\nmt_0| < 10^{-22}\,\textrm{GeV}\,,\nn\\
|\xx	5	9	\nmo_0| < 10^{-22}\,, && 
|\xx	5	{10}	\nmt_0| < 10^{-22}\,,\nn\\
|\xx	6	2	\sym	000|<10^{-22}\,\textrm{GeV}^{-1}\,. &&
\eea
The above limits represent the primary result of this work. 
To our knowledge,
they provide the first measurement of $\ze^{(6)}_2 S_{000}$ 
as well as the first measurement of any nonmetricity component 
inside matter.

The measurement~\rf{measurement} 
performed at the NG-6 slow-neutron beamline 
at NIST's Center for Neutron Research
has already appeared in the literature~\cite{Nico05}. 
Neutrons with transverse spin polarization
traversed 1~meter of liquid $^{4}$He
that was kept at a temperature of $4\,$K
in a magnetically shielded cryogenic target.
The neutron beam's energy distribution 
was well approximated by a Maxwellian 
with a maximum close to $3\,$meV. 
Paralleling the usual light-optics set-up 
of a crossed polarizer--analyzer pair,
the experiment searched for 
a nonzero rotation in the neutrons' polarization.
Further details of this measurement 
can be found in Refs.~\cite{Bas09, Sno11, Micherdzinska11, Swa10, Sno12, Sno15}.
The result quoted in the above Eq.~\rf{measurement}
represents the upper limit on 
the parity-odd neutron-spin rotation angle per unit length 
in liquid helium at $4\,$K
extracted from the measured data.

The usual SM
incorporates known parity-violating physics 
that can also lead to neutron spin rotation, 
for instance via interactions with electrons or nucleons. 
In~fact,
this phenomenon has been measured 
in heavy nuclei~\cite{Forte80, Heckel82, Heckel84}.
A convincing interpretation of the above nonmetricity constraint
therefore requires a discussion 
of this SM background.
From a theoretical perspective, 
parity violation in neutron--electron physics
in the SM 
is well understood. 
In particular, 
it is suppressed relative to the parity-odd neutron--nucleon interaction
by the weak charge $(1-4 \sin^{2}\theta_{W}) \approx 0.1$.
The neutron--nucleon parity violation,
on the other hand,
is induced by quark--quark weak interactions.
This system also involves 
the strong-coupling limit of QCD,
which still evades solid theoretical tractability.
Nevertheless,
nucleon--nucleon weak-interaction amplitudes
have been argued
to be six to seven orders of magnitude below 
strong-interaction amplitudes 
at neutron energies 
relevant for our present purposes~\cite{Sto74}.
Although reliant on phenomenological input 
in the form of 
nuclear parity-violation data folded into a specific model,
the value $d\phi_{PV}/dL=-6.5\pm2.2\times 10^{-7}\,$rad/m
for the SM spin rotation in the neutron--$^{4}$He system
is regarded as the most decent theoretical estimate~\cite{Des98}. 
Our experimental upper limit on nonmetricity~\rf{bound1}
is larger than this SM-background estimate. 
For this reason,
we disregard
the remote possibility of a cancellation
between SM and nonmetricity contributions 
to neutron spin rotation.

To determine additional limits on in-matter nonmetricity,
one could also consider using data
from other high-precision parity-violation experiments.
One example in the context of neutrons
are measurements of parity-breaking effects in atoms 
that are affected by the nuclear anapole moment 
and arise from parity-odd interactions 
between nucleons~\cite{Zel57, Woo97}.
An idea for 
extracting nonmetricity constraints 
involving electrons 
could,
for example,
be based on
the consistency between 
the theoretical SM result 
and the experimental value 
of the weak charge of the $^{133}$Cs atom~\cite{Bouchait:2005}.

Additional nonmetricity components 
may become experimentally accessible
with a set-up 
in which both the slow-neutron beam 
as well as the nuclear target 
are polarized:
the aligned target spins 
would coherently generate
large-scale anisotropic components of $N_{\al\be\ga}$, 
which were disregarded in our above analysis.
High-sensitivity studies 
of this type
have received considerable attention
for quite some time~\cite{Bar12}.
The neutron--nucleus scattering amplitude 
exhibits a significant polarization dependence,
an effect known as nuclear pseudomagnetic precession~\cite{Bar65}:
the neutron's spin precesses about the nuclear polarization vector 
as the neutron traverses the polarized medium. 
In the past, 
this method has been employed 
to determine the spin dependence 
of neutron--nucleus scattering cross sections 
for a number of nuclei~\cite{Abr72}.
However,
the nuclear-pseudomagnetism spin-precession contributions
from the strong neutron--nucleus interaction
to such a measurement
are substantial
and currently evade theoretical treatment 
from first principles.
It is therefore expected that
the experimental reach 
regarding in-matter anisotropic $N_{\al\be\ga}$ components 
would be more modest
than that in this study.

We finally mention that 
a high-precision transmission-asymmetry measurement
utilizing transversely polarized $5.9\,$MeV neutrons 
was performed in a nuclear spin-aligned target of holmium~\cite{Huffman97}.
This experiment 
explored the presence of
P-invariant but T-violating 
interactions of the neutron.
The measurement yielded 
$A_{5}={\sigma_{P} \over \sigma_{0}}=+8.6\pm 7.7\textrm{(stat.+sys.)}\times10^{-6}$. 
Here,
$A_{5}$ denotes the transmission asymmetry for neutrons 
polarized parallel and antiparallel 
to the normal of the plane spanned by
the neutron momentum 
and the spin polarization of the holmium target.
An open question is whether or not 
polarized nuclear matter 
generates an effective $N_{\al\be\ga}$ 
that differs from that of unpolarized nuclear matter, 
and how such a difference would manifest itself
in this experiment.
That said, 
the neutron energy in this measurement
remains nonrelativistic, 
so our above methodology 
should continue to be applicable.

\section{Summary}
\label{sum}

In this work,
we have considered the possibility 
of nontrivial nonmetricity in nature.
We have argued that 
in this context 
an effective nonmetricity field
could be generated
inside a liquid ${}^4$He target. 
We have shown that
the spin of nonrelativistic neutrons 
traversing such a target 
would then precess.
This prediction, 
together with existing data on neutron spin rotation 
in liquid ${}^4$He,
implies the primary result of this work,
namely the bound~\rf{bound1}. 
To our knowledge,
this is the first experimental limit 
on in-matter nonmetricity.

We have further concluded that
it would be difficult to improve our bound
via higher-precision spin-rotation data 
due to the conventional SM background
arising from quark--quark weak interactions.
However, 
other atomic and nuclear parity-violation tests
may have the potential 
to yield complementary limits 
on nonmetricity interactions of neutrons and electrons.
Moreover,
polarized slow-neutron transmission measurements
through polarized nuclear targets 
could be studied 
with the approach presented in this work
and may give bounds on additional in-matter $N_{\al\be\ga}$ components.
We encourage other researchers 
to perform further nonmetricity searches 
using the general framework 
employed in this study
with the aim to turn nonmetricity tests
into a more quantitative experimental science.

\section*{Acknowledgments}

This work was supported by the DOE, 
by the NSF under grant No.~PHY-1614545, 
by the IU Center for Spacetime Symmetries, 
by the IU  Collaborative Research and Creative Activity Fund of the Office of the Vice President for Research, 
by the IU Collaborative Research Grants program,
by the National Science Foundation of China 
under grant No.~11605056, 
and by the Chinese Scholarship Council.
RL acknowledges support from the Alexander von Humboldt Foundation.
ZX is grateful for the hospitality of the IUCSS,
where most of this work was completed.


\end{document}